\documentclass[twocolumn,amsmath,amssymb,floatfix,prb,showpacs,footinbib]{revtex4}
\usepackage{amsmath,amssymb,natbib,bm,graphicx,url,psfrag}

\usepackage{color,times}

\newcommand{\NN}{\mathbb{N}}

\newcommand{\vc}[1]{\mathbf{#1}}

\newcommand{\abs}[1]{\left|#1\right|}
\newcommand{\bra}[1]{\left\langle \, #1 \,\right|}
\newcommand{\ket}[1]{\left|\, #1 \, \right\rangle}

\newcommand{\be}{\begin{equation}}
\newcommand{\ee}{\end{equation}}

\begin{document}

\title{Superfluid--Mott Insulator Transition of Light in the Jaynes-Cummings Lattice}
\author{Jens Koch}
\affiliation{Departments of Physics and Applied Physics, Yale University, PO Box 208120, New Haven, CT 06520, USA}
\author{Karyn Le Hur}
\affiliation{Departments of Physics and Applied Physics, Yale University, PO Box 208120, New Haven, CT 06520, USA}

\begin{abstract}
Regular arrays of electromagnetic resonators, in turn coupled coherently to
individual quantum two-level systems, exhibit a quantum phase transition of
polaritons from a superfluid phase to a Mott-insulating phase. The critical behavior of such a Jaynes-Cummings lattice thus resembles the physics of the Bose-Hubbard model. 
We explore this analogy by elaborating on the mean-field theory of the phase transition, and by presenting several useful mappings which pinpoint both similarities and differences of the two models. We show that a field-theory approach can be applied to prove the
existence of multicritical curves analogous to the multicritical points of the Bose-Hubbard model, and we provide analytical expressions for the position of these curves.
\end{abstract}

\date{August 18, 2009} 
\maketitle

\section{Introduction\label{sec:Introduction}}
Quantum phase transitions in interacting systems composed of particles of more than one species have recently produced a lot of interest. A host of problems in condensed matter physics fall into this category, prime examples being heavy fermions in Kondo lattices,\cite{v._lhneysen_fermi-liquid_2007} multi-component systems of ultracold atoms in optical lattices,\cite{altman_phase_2003,han_quantum_2004,orth_dissipative_2008} and the ensemble of two-level atoms interacting with a bosonic mode described by the Dicke model.\cite{dicke_coherence_1954,emary_quantum_2003}

\begin{figure}
\centering    
		\includegraphics[width=1.0\columnwidth]{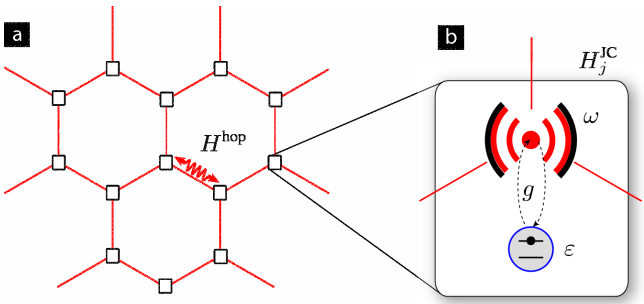}
	\caption{(Color online) Schematic of the Jaynes-Cummings lattice system, consisting of an array of electromagnetic resonators, see (a), with a coupling between nearest-neighbor lattice sites due to photon hopping. Each resonator is coherently coupled to a two-level system, shown in (b). \label{fig:array}}
\end{figure}

A recent addition to this list of systems is the Jaynes-Cummings (JC) lattice model, see Fig.\ \ref{fig:array}. It describes an array of electromagnetic resonators, each coupled coherently to a single quantum mechanical two-level system (also referred to as ``atom" or ``qubit"). Recently, this JC coupling\cite{jaynes_comparison_1963}  has been studied extensively in the context of quantum computation and quantum optics.\cite{raimond_colloquium:_2001,schoelkopf_wiring_2008}
The interplay of Jaynes-Cummings interaction, observable in experiments as a photon blockade,\cite{birnbaum_photon_2005}
and the transfer of photons between nearest-neighbor resonators within the array render this problem nontrivial, and open a new route to study
strongly correlated photon physics and quantum phase transitions of light.\cite{hartmann-np,hartmann-njp,lukin,makin} Similar to ultracold atom systems, the Jaynes-Cummings lattice could
provide a novel quantum simulator of condensed matter Hamiltonians.

Previous work on the JC lattice model has so far explored several aspects of its theory, and has provided the first valuable hints towards possible realizations of the model in experiments. The early mean-field treatment\cite{greentree_quantum_2006,na_strongly_2008} has provided evidence for a phase transition from  Mott-insulating phases to a superfluid phase of polaritons, resembling in large parts the physics known from the Bose-Hubbard model.\cite{fisher_boson_1989} Both numerical and analytical methods have subsequently been employed to extract the phase boundaries beyond the mean-field approximation,\cite{aichhorn_quantum_2008,zhao_insulator_2008,rossini_mott-insulating_2007,hohenadler,schmidt_strong_2009} and to study the critical behavior,\cite{zhao_insulator_2008,hohenadler,schmidt_strong_2009} the spectrum of excitations,\cite{aichhorn_quantum_2008,hohenadler,schmidt_strong_2009} and the physics in the limit of only a few coupled cavities.\cite{angelakis_photon-blockade-induced_2007} For a more comprehensive review of the relatively young field, we refer the reader to the excellent paper by Hartmann et al.\cite{hartmann_quantum_2008-1}

The prime motive of the present paper is to clarify the nature of the correspondence between Bose-Hubbard model and Jaynes-Cummings lattice model. After the introduction of the specific model in Section \ref{sec:model}, we provide a systematic discussion of the asymptotic behavior of the JC lattice model in several useful limits (Section \ref{sec:asymptotic}). The knowledge of these asymptotics aides our discussion in the subsequent three main sections of our paper. The mean-field treatment is revisited in Section \ref{sec:meanfield}, where we provide analytical expressions for the phase boundary for general atom-photon detuning, and elaborate on several issues with previous mean-field treatments. The correspondence beyond the mean-field level is explored in Sections \ref{sec:mapping} and \ref{sec:fieldtheory}, where we discuss several useful mappings of the model and formulate the field-theoretic treatment of the Jaynes-Cummings lattice model. The latter approach allows the existence proof of multicritical curves analogous to the multicritical points present in the critical behavior of the Bose-Hubbard model. We conclude our paper with a summary and outlook of relevant future research directions in Section \ref{sec:conclusions}.

\section{Jaynes-Cummings lattice model\label{sec:model} }
The Jaynes-Cummings lattice model is described by the Hamiltonian ($\hbar=1$)
\begin{equation}\label{HJC1}
H= \sum_j H^\text{JC}_j +H^\text{hop}-\mu N,
\end{equation}
where 
\begin{equation}\label{HJC2}
H^\text{JC}_j =  \omega a^\dag_ja_j + \varepsilon\sigma^+_j\sigma^-_j 
+g(a^\dag_j\sigma^-_j + \sigma^+_ja_j)
  \end{equation}
denotes the Jaynes-Cummings Hamiltonian on site $j$. (A succinct summary of this model, exactly solvable for one site and well-known in quantum optics, is given in Appendix \ref{app:JC}.) The operator $a_j$ ($a_j^\dag$) annihilates (creates) a photon in the resonator at site $j$, in a mode with frequency $\omega$. Each resonator hosts a single two-level atom, whose transition frequency is given by $\varepsilon$. Adopting the (pseudo-)spin language, the corresponding lowering and raising operators are denoted by the Pauli matrices $\sigma_j^\pm$. The atom-photon coupling of strength $g$ allows for the coherent interconversion between photonic and atomic excitations within each resonator.
We note that for most systems, the JC Hamiltonian \eqref{HJC2} is only obtained when invoking the rotating wave approximation (RWA). 
The contributions from counter-rotating terms become significant when the coupling strength $g$ is of the same order as the resonator or atom frequency -- a fact neglected in several recent publications where $g=\varepsilon$ was used.\cite{zhao_insulator_2008,aichhorn_quantum_2008} 

The hopping of photons between nearest-neighbor cavities is described by
\begin{equation}\label{Hhop}
H^\text{hop}= - \kappa \sum_{\langle i,j\rangle} (a_i^\dag a_j + a_j^\dag a_i),
\end{equation}
where the hopping strength is parameterized by $\kappa$, so that the overall lifetime of each resonator is given by $(z_c\kappa)^{-1}$, $z_c$ being the coordination number of the lattice.

The chemical potential term in Eq.\ \eqref{HJC1} is in accordance with our treatment of the system within the grandcanonical ensemble. The chemical potential $\mu$ couples to the (conserved) number of polaritons, $N=\sum_j(a_j^\dag a_j+\sigma_j^+\sigma_j^- )$, and as usual plays the role of a Lagrange multiplier fixing the mean total number of polaritons on the lattice, $\langle N\rangle$. By contrast to the situation in ultracold atom systems, it is important to emphasize that, for the physical realizations of the JC lattice system proposed so far,\cite{hartmann_quantum_2008-1} the chemical potential $\mu$ is not a directly accessible ``knob." Instead, appropriate preparation schemes have to be devised to access states with different mean polariton numbers, see e.g.\ Ref.\ \onlinecite{angelakis_photon-blockade-induced_2007}.

\section{Asymptotic behavior of the JC lattice model\label{sec:asymptotic}}
Similar to the situation encountered in the study of the Bose-Hubbard model,\cite{fisher_boson_1989} the JC lattice model is not generally amenable to an exact solution. In the following, we discuss two particularly simple limits: the atomic limit and the hopping-dominated limit where $\kappa/g\ll1$ and $\kappa/g\gg1$, respectively. The understanding of these limits serves as a useful building block and sets the stage for our subsequent study of the more intricate and interesting regimes of the JC lattice model.

\subsection{Atomic limit: $\kappa/g\ll1$}
In this limit, the photon hopping  $H^\text{hop}$ may be treated perturbatively. To leading order, the Hamiltonian decouples in the site index and reduces to pure Jaynes-Cummings physics, described by the  Hamiltonian $H^\text{JC}_j-\mu n_j$ and identical for each site $j$.  Consequently, the ground state of the lattice system corresponds to a  product state $\ket{\Psi}^{\otimes_j}=\ket{\Psi}_{j=1}\otimes\ket{\Psi}_2\otimes\cdots$ of the ground states of each $H^\text{JC}_j$. (To simplify notation, we drop the site index $j$ for the remainder of this section.)

 The eigenstates of the JC Hamiltonian correspond to polaritons -- quasiparticles consisting of both photonic and atomic excitations. Their spectrum, shifted by the chemical potential according to $E^\mu_0=E_0-\mu n$ and $E^\mu_{n\pm}=E_{n\pm}-\mu n$, composes the usual  ``dressed-state ladder" of the JC model: $E_0=0$ and 
\begin{equation}\label{JCenergies}
E_{n\pm}=n\omega+\Delta/2\pm [(\Delta/2)^2+n g^2]^{1/2} \quad (n\ge1).
\end{equation}
 (For details, see Appendix \ref{app:JC}.) Here and in the following, $\Delta\equiv\varepsilon-\omega$ denotes the detuning between the two-level atom and the resonator frequency. The polariton states $\ket{n\pm}$ are simultaneous eigenstates of $H^\text{JC}$ and of the polariton number $n=\sigma^+\sigma^- +a^\dag a$. (Note that while the symbol ``$N$" refers to the polariton number on the entire lattice, ``$n$" denotes the polariton number on a single site.) Each parity doublet $\pm$ is adiabatically connected to the symmetric and antisymmetric superposition of atomic and photonic excitation, $\ket{n\pm}=(\ket{n\downarrow}\pm\ket{(n-1)\uparrow})/\sqrt{2}$, which form the eigenstates in the resonant case, $\Delta=0$. 

\begin{figure*}
	\centering
		\includegraphics[width=0.95\textwidth]{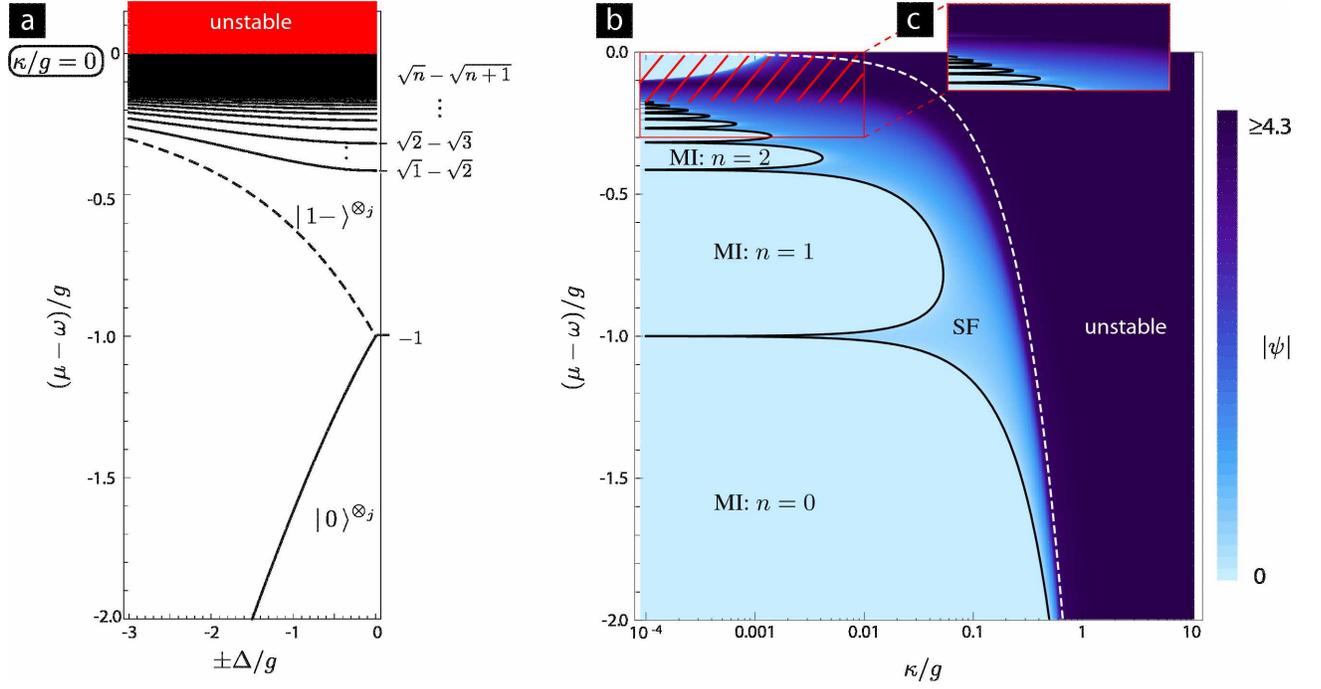}
		\caption{(Color online) Properties of the JC lattice model in the resonant case, $\Delta=0$. \textbf{(a)} Ground states of the JC lattice system in the atomic limit, $\kappa=0$, as a function of detuning $\Delta$. The ground state of the system is given by a product state of Jaynes-Cummings eigenstates on each lattice site $j$. Depending on the chemical potential, the system assumes either the state $\ket{0}^{\otimes_j}$ or one of the antisymmetric states $\ket{n-}^{\otimes_j}$. Degeneracies between $\ket{n-}$ and $\ket{(n+1)-}$  mark the onset of superfluidity, occurring for finite photon hopping, $\kappa>0$. The onset points, for zero detuning located at $(\mu-\omega)/g=\sqrt{n}-\sqrt{n+1}$, become dense as $\mu$ approaches $\omega$. For $\mu>\omega$, the system becomes unstable. Degeneracies occur at the same chemical potential for negative and positive detunings $\Delta$ (solid curves), except for the lowest degeneracy between $\ket{0}^{\otimes_j}$ and $\ket{1-}^{\otimes_j}$ where the $\Delta>0$ case is given by the dashed curve.
\textbf{(b)} Mean-field phase diagram of the resonant JC lattice system as a function of the effective chemical potential, $\mu-\omega$, and photon hopping strength $\kappa$. The color scale shows the magnitude of the order parameter $\psi=\langle a \rangle$.  The value of $\psi$ reveals the Mott-insulating phases (denoted ``MI") with $\psi=0$ and fixed number $n=0,1,\ldots$ of polaritons per site, and the superfluid phase (``SF") with $\psi>0$. The phase boundary can be obtained analytically, cf.\ Eq.\ \eqref{pboundary}, and is marked by a black curve. For sufficiently large photon hopping strength, the system becomes unstable with respect to addition of polaritons. A crude estimate of the onset of instability is given by $(\mu-\omega)=z_c\kappa$, depicted by the white dashed curve. In the hatched region close to $\mu-\omega=0$, numerical results are unreliable when using a fixed cutoff for the maximum photon number. (c) Improved numerical results near $\mu-\omega=0$ [range of $\mu-\omega$ and $\kappa$ as marked by the rectangle in panel (b)] can be obtained by employing a ``sliding" truncation (see text) centered at the photon occupation number  obtained in the atomic limit.\label{fig:phase-diagram}}\end{figure*}

To specify the ground state of the system for given chemical potential $\mu$, we must find the state $\ket{n\alpha}$ with minimal energy, i.e.\
\be
E_{n\alpha}^\mu = \min \{E_0^\mu,E_{1\pm}^\mu,E_{2\pm}^\mu,\ldots\}.
\ee
Immediately, one can rule out all symmetric states $\ket{n+}$ as ground states, since $E^\mu_{n+}\ge E_{n-}^\mu$. Hence, the ground state is either the $0$-polariton state $\ket{0}$ or one of the antisymmetric states $\ket{n-}$. To further specify the ground state, first consider the limit $(\omega-\mu)\gg g,\abs{\Delta}$. (Both $\omega$ and $g$ are assumed to be non-negative.) In this case, the first term of Eq.\ \eqref{JCenergies} dominates over the second term, and the lowest energy is reached for the zero-excitation state $\ket{0}$. By decreasing the quantity $(\omega-\mu)$, one reaches a point where admitting an excitation to the system becomes energetically favorable. This happens precisely when $E_0^\mu=E_{1-}^\mu$. Repeating this argument, one finds an entire set of such degeneracy points, determined by $E_{n-}^\mu=E_{(n+1)-}^\mu$. The full set of degeneracy points is given by 
\begin{equation}
\label{b2}
(\mu-\omega)/g = \sqrt{n+(\Delta/2g)^2}-\sqrt{n+1+(\Delta/2g)^2}, 
\end{equation}
where $\quad n=0,1,\ldots$, which reduces to $(\mu-\omega)/g=\sqrt{n}-\sqrt{n+1}$ in the resonant case. Our results for the atomic limit, depicted in Fig.\ \ref{fig:phase-diagram}(a), are consistent with results previously presented by Greentree and coauthors.\cite{greentree_quantum_2006}

As shown in Fig.\ \ref{fig:phase-diagram}(a), the spacing between degeneracy points becomes closer as $(\mu-\omega)$ is increased towards zero. More precisely, by inspection of Eq.\ \eqref{b2} one finds that for $n\to\infty$, $(\mu-\omega)$ tends to zero, independent of the detuning $\Delta$. In other words, the point $(\mu-\omega)=0$ where the chemical potential is tuned to exactly compensate the resonator frequency, plays the role of an accumulation point. In Section \ref{sec:meanfield} below, we will demonstrate that the existence of this accumulation point necessitates extra care in the numerical analysis when approaching $(\mu-\omega)=0$.

This leaves the question of what happens for $(\mu-\omega)>0$. In this case, the system becomes unstable: adding more polaritons to the system always \emph{lowers} the total energy. This behavior, if not accompanied by an additional mechanism that would ultimately limit the polariton number, is unphysical. Hence, we restrict ourselves to values of $\mu\in[-\infty,\omega]$ in the following. We will encounter a similar instability in the following section describing the limit of large photon hopping.

\subsection{Hopping-dominated limit: $\kappa/g\gg1$\label{sec:hopdom}}
For large photon hopping, $\kappa/g\gg1$, the Hamiltonian $H^\text{hop}$  overwhelms the atom-photon coupling, and the latter may be treated perturbatively.  As a crude approximation, we consider the order $g^0$, i.e.\ atom-photon coupling is dropped and the atomic and photonic systems decouple completely. The ground state energy then consists of the atomic contribution (all atoms occupying their respective ground states) and the photonic contribution from the  bosonic tight-binding model
\be
H^\text{tb}=(\omega-\mu)\sum_i a_i^\dag a_i - \kappa \sum_{\langle i,j \rangle}\left(a_i^\dag a_j + a_j^\dag a_i\right).
\ee
As usual, the tight-binding model may be diagonalized in terms of single-particle Bloch waves. For the relevant cases of a 2d cubic lattice and a 2d honeycomb lattice, the resulting energy dispersions are given by
\be
\epsilon(\vc{k}) = (\omega-\mu) -2\kappa \sum_{i=x,y} \cos(k_i a)
\ee
for the cubic lattice,\cite{ashcroft_solid_1976} and
\be
\epsilon_\pm(\vc{k}) = (\omega-\mu)\pm\kappa\abs{1+e^{-ik_xa}+e^{-i(k_x-k_y)a}} 
\ee
for the honeycomb lattice ($a$ denotes the lattice constant).\cite{wallace_band_1947} In the energy dispersion of the honeycomb lattice, the two signs refer to the lower and upper ($\pi$ and $\pi^*$) band. Independent of the specific lattice type, the bosonic ground state is obtained by $N$-fold occupation of the  $\vc{k}=0$ ($\pi$) state with corresponding energy 
\be\label{gstate}
E_0 = N(\omega-\mu) - Nz_c\kappa.
\ee
This result reveals the presence of an instability at large photon hopping strength: for $z_c\kappa > \omega-\mu$ the ground state energy becomes negative and can be made arbitrarily small by increasing the polariton number $N$. (We note that a similar instability formally exists for the Bose-Hubbard model in the region of \emph{negative} chemical potential.) This instability is already present in a mean-field treatment of the JC lattice system,  where its presence has apparently been overlooked in Refs.\ \onlinecite{greentree_quantum_2006} and \onlinecite{lei_quantum_2008}.

\subsection{Effective Hubbard-$U$ in the dispersive regime?\label{sec:effU}}
A prime motive of the present paper is the analysis of differences and similarities between the Bose-Hubbard model and the Jaynes-Cummings lattice model. As we shall demonstrate, based on mean-field theory, exact and approximate mappings, as well as field theoretical methods, there exist strong parallels between the two models. We find, however, that the direct comparison of the two models in terms of an ``effective Hubbard-$U$" is generally not appropriate. Correcting for a missing factor of $1/n$ in Ref.\ \onlinecite{greentree_quantum_2006}, one may attempt to define such an effective Hubbard-$U$ via the relation
\be\label{U}
U_{n\pm}=\left( E_{(n+1)\pm}^\mu-E_{n\pm}^\mu+\mu \right)/n.
\ee
As argued by Greentree et al., this effective $U$ approaches zero in the limit of large polariton numbers $n$ per resonator -- consistent with the predominance of the superfluid phase in this regime. By contrast, we find that the system does \emph{not} approach a constant Hubbard-$U$ in the large detuning limit (as stated in Ref.\ \onlinecite{greentree_quantum_2006}) when using the correct expression \eqref{U}. To see this, we consider the dispersive regime $g/\Delta\ll1$ with polariton numbers below the critical photon number, $n\ll n_\text{c}=\Delta^2/4g^2$.
In this case, the dressed-state ladder decomposes into two nearly harmonic ladders shifted by the detuning $\Delta$,
\be
E_{n\pm}^\mu \simeq n(\omega-\mu) +\frac{\Delta}{2} \pm \frac{\abs{\Delta}}{2},
\ee 
i.e.\ photons and atoms are effectively decoupled. This situation is not correctly captured by a Bose-Hubbard model, and indeed we find that the effective interaction $U_{n\pm}=\omega/n$ would need to be strongly polariton-number dependent in this case.


\section{Mean-Field theory\label{sec:meanfield}}

The mean-field theory (MFT) of the Jaynes-Cummings lattice model has previously been discussed by Greentree et al.\cite{greentree_quantum_2006} It reveals a second-order phase transition from a superfluid phase of polaritons to Mott-insulating phases. We revisit this treatment, pointing out several important issues not addressed in Ref.\ \onlinecite{greentree_quantum_2006}.

Starting from the JC lattice Hamiltonian, Eqs.\ \eqref{HJC1}--\eqref{Hhop}, the MFT is constructed by a decoupling of the lattice sites. We apply the general prescription of replacing an interaction term involving two operators $A$ and $B$ by its mean-field expression, $AB\to\langle A\rangle B + A\langle B\rangle - \langle A\rangle\langle B\rangle$, to the photon-hopping term  in the Jaynes-Cummings lattice model. This yields
\begin{align}
H^\text{hop}&=\kappa \sum_{\langle i,j\rangle} (a_i^\dag a_j + a_j^\dag a_i)=\kappa \sum_i\sum_{j\in \mathrm{nn}(i)} a_i^\dag a_j\\\nonumber
&\to
\kappa \sum_i\sum_{j\in \mathrm{nn}(i)} \left[ \langle a_i^\dag \rangle a_j +a_i^\dag \langle a_j \rangle  -  \langle a_i^\dag \rangle  \langle a_j \rangle \right],
\end{align}
where $\mathrm{nn}(i)$ denotes the set of sites that are nearest neighbors to site $i$.
Translation invariance implies that the expectation values are actually site-independent. Introducing the order parameter as $\psi=z_c\kappa\langle a_i\rangle$,\footnote{We note that for numerical purposes it is convenient to carry out the minimization with respect to $\bar{\psi}=\kappa\psi$ in order to avoid numerical problems in the limit of $\kappa\to0$.} the mean-field Hamiltonian reads $H^\text{mf}=\sum_j h^\text{mf}_j$ with
\begin{align}\label{hmf1}
&h^\text{mf}_j =  \frac{1}{2}(\varepsilon-\mu)\sigma_j^z + (\omega-\mu) a_j^\dag a_j +  g(a_j^\dag \sigma_j^- + \sigma_j^+ a_j) \nonumber\\
&\qquad   - (a_j \psi^* + a_j^\dag \psi) +\frac{1}{z_c\kappa}\abs{\psi}^2.
\end{align}
Here, $z_c$ is the coordination number of the lattice. It is useful to note that the dimension and precise geometry of the spatial lattice underlying the JC lattice model enter the mean-field theory only through $z_c$. Moreover, a change of the coordination number is equivalent to a rescaling of the photon hopping strength. Specifically, within MFT the substitution $z_c\to r z_c$ has the same effect as the rescaling $\kappa \to r\kappa$ while keeping $z_c$ fixed.

Throughout this paper, we focus on the  quantum phase transition of the JC lattice model at zero temperature. In this case, the order parameter is determined by minimization of the ground state energy $E_0(\psi)$ of the mean-field Hamiltonian $h^\text{mf}$. Figure \ref{fig:phase-diagram}(b) presents numerical results from this analysis. Consistent with numerical results previously published by Greentree et al.,\cite{greentree_quantum_2006} our phase diagram shows lobes where the order parameter $\psi$ vanishes. These lobes, which are the analogs of the Mott-insulator (MI) phases in the Bose-Hubbard model,\cite{fisher_boson_1989}  border on a superfluid phase with nonzero order parameter, $\abs{\psi}>0$. The onset of superfluidity is continuous, in line with the assumption of a second-order phase transition.

In addition to the Mott-insulating and superfluid phases discussed by Greentree et al., our results also reveal the onset of the polariton-number instability expected for the hopping-dominated regime, see Section \ref{sec:hopdom} above. The unstable region can be identified by the darkest coloring in Fig.\ \ref{fig:phase-diagram}(b). In the numerical analysis of the mean-field equations, the instability manifests itself in solutions with  excessively large values of the order parameter $\psi$. More specifically, in the unstable region, the value of the order parameter is entirely determined by the photon cutoff, i.e.\ the truncation of the photon Hilbert space necessary in the numerical treatment. When truncating this Hilbert space at some maximum photon number $n_\text{max}$, one can confirm that the order parameter obeys the relation $\psi\sim n_\text{max}^{1/2}$ in the unstable region. This instability appears to be present in the numerical results obtained by Greentree et al.,\cite{greentree_quantum_2006} and very likely also in qualitatively similar results obtained by Lei and Lee for the Dicke-Bose-Hubbard model.\cite{lei_quantum_2008} In both references, the instability is not mentioned, and it has not been discussed elsewhere to the best of our knowledge. A rough estimate for the border of the instability region may be obtained from the hopping-dominated limit by setting the ground state energy, Eq.\ \eqref{gstate}, to zero. This leads to the relation $(\mu-\omega)=-z_c\kappa$, shown as the dashed curve in Fig.\ \ref{fig:phase-diagram}(b), which is in reasonable agreement with the numerical results.

There exists a second region in the phase diagram  which is highly sensitive to the specific photon truncation scheme. This region close to $\mu-\omega=0$, marked by hatching in Fig.\ \ref{fig:phase-diagram}(b), appears to suggest the emergence of a large Mott-insulating phase reaching from $\mu-\omega\simeq-0.1$ all the way into the $\mu-\omega>0$ region. This feature, present also in all mean-field phase diagrams of References \onlinecite{greentree_quantum_2006} and \onlinecite{lei_quantum_2008}, turns out to be a pure artefact of the truncation of Hilbert space at some maximum photon number. Indeed, its presence would be rather difficult to reconcile with the atomic-limit prediction of an infinite number of MI phases with $\mu-\omega=0$ being an accumulation point and with the region $\mu-\omega>0$ being unstable. 

To illustrate this point, we have employed an alternative ``sliding" truncation scheme based on the number of excitations in the atomic limit. Specifically, if the ground state in the atomic limit is given by $\ket{n-}^{\otimes_j}$, we take into account a total of $2\Delta_n$ photon Fock states, with numbers $n'$ within
\be
\max\{0,n-\Delta_n \}\le n'\le n+\Delta_n.
\ee
The results from this sliding truncation scheme, presented in Fig.\ \ref{fig:phase-diagram}(c), uncover that results from the maximum photon number truncation scheme are unreliable in the hatched region. As expected from the atomic limit, the sliding truncation scheme shows no large Mott-insulating region close to $\mu-\omega=0$, but only the sequence of ever smaller MI phases approaching this chemical potential value.
 
The phase boundary between MI and superfluid phases can be determined in a way similar to the procedure for the Bose-Hubbard model.\cite{fisher_boson_1989,sachdev_quantum_2000} 
In the critical region, the superfluid order parameter $\psi$ is small and the terms in $h^\text{mf}$ involving $\psi$ may be treated perturbatively. This leads to an expansion of the ground state energy in powers of $\psi$,
\begin{equation}\label{mfexp}
E_0(\psi)=E_0^\text{mf}+r\abs{\psi}^2+\frac{1}{2}u\abs{\psi}^4+\mathcal{O}(\abs{\psi}^6).
\end{equation}
Note that the invariance of $h^\text{mf}$ under the global gauge transformation
\be
\psi\to\psi e^{i\phi},\quad a_j\to a_j e^{-i\phi}, \quad \sigma_j^-\to\sigma_j^- e^{-i\phi},
\ee
implies that all odd-order terms in the expansion \eqref{mfexp} vanish. 

Eq.\ \eqref{mfexp} represents the standard situation of a quadratic plus quartic potential, ubiquitous in the study of mean-field phase transitions. 
The resulting mean-field critical exponents are $\beta=1/2$ for the critical exponent of the order parameter, $\psi\sim\bar{\mu}^{\beta}$, and $\alpha=0$ for the exponent of the compressibility, defined by $\kappa_s\sim \bar{\mu}^{-\alpha}$.\cite{herbut_modern_2007} Here, $\bar\mu=(\mu-\mu_c)/\mu_c$ measures the relative chemical potential difference from the critical value. Other critical exponents can be established via scaling and hyperscaling relations, and we discuss the role of the dynamical critical exponent $z$ in Section \ref{sec:fieldtheory}.

\begin{figure}
	\centering
		\includegraphics[width=1.0\columnwidth]{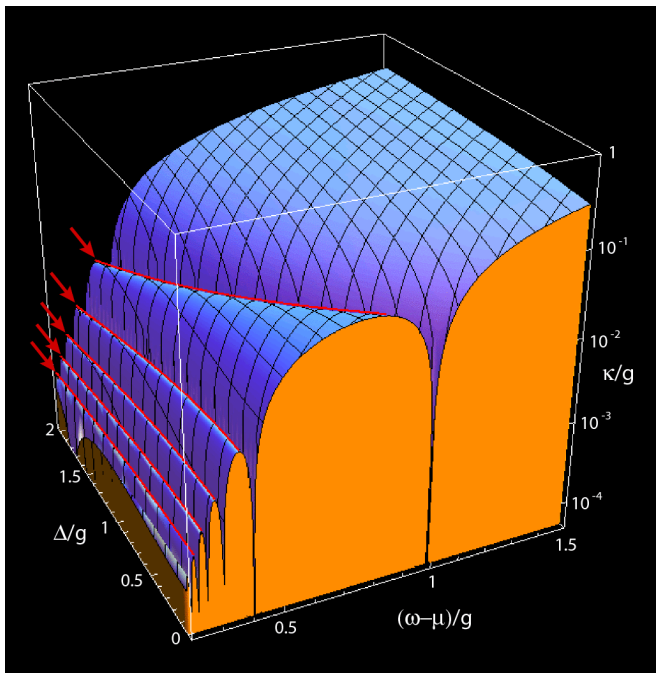}
	\caption{(Color online) Mean-field phase boundary as a function of effective chemical potential $(\mu-\omega)$ and atom-resonator detuning $\Delta$.\label{fig:3dphase} The cross-section shows the Mott-insulating lobes at zero detuning (in orange/light gray). With exception of the $\ket{0}$ lobe, all lobes become narrower when increasing the detuning $\Delta$. Arrows and curves in red/dark gray mark the positions of multicritical lines where the system switches from the universality class of the generic superfluid--Mott insulator transition to the universality class of the soft-spin $(d+1)$--dimensional XY model.}
\end{figure}

As usual, the phase boundary is specified by the condition $r=0$ (assuming that $u>0$).
The quadratic expansion coefficient $r$ can be expressed as $r= R_n + (z_c\kappa)^{-1}$
where $R_n$ can be obtained from second-order perturbation theory in the photon hopping. In the resonant case ($\Delta=0$), the contributions are given by
\begin{align}\label{Req}
 R_{n=0} &=\frac{1}{2}\sum_{\alpha=-1,+1}  \frac{1}{\mu-\omega+\alpha g},\\
R_{n>0} &= \frac{1}{4} \sum_{\alpha=-1,+1} \bigg[ \frac{(\sqrt{n}+\alpha\sqrt{n-1})^2}{-(\mu-\omega)-(\sqrt{n}-\alpha\sqrt{n-1})g}\nonumber\\
&\qquad\qquad
+\frac{(\sqrt{n+1}+\alpha\sqrt{n})^2}{(\mu-\omega)-(\sqrt{n}-\alpha\sqrt{n+1})g}\bigg],
\end{align}
where $n=0,1,\ldots$ enumerates the sequence of Mott-insulating lobes. For nonzero detunings, the lengthier expressions for the quantities $R_n$ are relegated to Appendix \ref{app:pert}. The full phase boundary is then obtained analytically as the set of curves defined by 
\be
\kappa_n = \frac{1}{z_cR_n},\label{pboundary}
\ee
shown for the resonant case as black curves in Fig.\ \ref{fig:phase-diagram}(b), and for finite detunings in Fig.\ \ref{fig:3dphase}. We will return to the question of phase boundaries in Section \ref{sec:fieldtheory}, where the presence of multicritical lines will be discussed.

\section{Useful mappings for the JC lattice model\label{sec:mapping}}
The qualitative similarity in the phase diagrams of the Jaynes-Cummings lattice model and the Bose-Hubbard model has been one of the main driving forces in recent studies of the JC lattice problem. Both models display a quantum phase transition from a superfluid phase to Mott-insulating phases. Yet, as discussed in Section \ref{sec:effU}, it is in general not possible to view the JC lattice model as a simple Bose-Hubbard model with an effective Hubbard interaction $U$. Instead, the composite system of interacting photons and atoms can be mapped to a two-component Bose-Hubbard model or a general polariton model which reduces to the spin-1/2 quantum--XX model in the vicinity of the degeneracy points between adjacent Mott lobes. We present the corresponding mappings and discuss their implications in the following three subsections.

\subsection{Mapping to a two-component Bose-Hubbard model}
As the first useful reformulation of the JC lattice model we explore its mapping to a two-component Bose-Hubbard model. For this mapping, the  annihilation and creation operators for spin waves are identified as $c_j=\sigma_j^-$, $c_j^\dag=\sigma_j^+$. As usual, these operators are neither bosonic nor fermionic: while on each individual site the $c_j$ obey anticommutation relations, $c_j$s from different sites commute,
\begin{align}
&\{c_j,c_j\}=0,\quad\{c_j,c_j^\dag\}=1\\
&[c_i,c_j]=[c_i,c_j^\dag]=0 \text{ for } i\not=j.
\end{align}
This deficiency may be cured by the Jordan-Wigner transformation\cite{jordan__1928} and its generalizations,\cite{fradkin_jordan-wigner_1989,batista_generalized_2001} which yield a proper fermionic description. Here, we instead employ the simple equivalence between spin-$1/2$ systems and hardcore bosons, valid for lattices in arbitrary dimension.\cite{friedberg_equivalence_1993} According to this equivalence, the energy spectrum of the spin-Hamiltonian $H(\sigma_j^+,\sigma_j^-)$ is identical to the set of finite eigenenergies of the hardcore boson model
\begin{equation}\label{hardcore}
H'=H(b_j^\dag,b_j)+\lim_{U\to\infty} \frac{U}{2}\sum_j b_j^\dag b_j(b_j^\dag b_j-1),
\end{equation}
where the first term is obtained from the Hamiltonian $H$ by replacing all spin operators with boson operators according to $\sigma_j^+\to b_j^\dag$, $\sigma_j^-\to b_j$. In our case, the resulting two-component Bose-Hubbard model reads
\begin{align}
H'=&\sum_j\bigg[ (\omega-\mu)a_j^\dag a_j + (\varepsilon-\mu)b_j^\dag b_j +\frac{U_\infty}{2}n_{bj}(n_{bj}-1) \nonumber\\
&+ g(a_j^\dag b_j + b_j^\dag a_j) \bigg] -\kappa \sum_{\langle i,j \rangle}\bigg[ a_i a_j^\dag + a_j a_i^\dag \bigg],\label{Uinf}
\end{align}
where $n_{bj}=b_j^\dag b_j$ is the number operator for the $b$ bosons on site $j$, and $U_\infty$ is a shorthand for the large-$U$ limit, cf.\ Eq.\ \eqref{hardcore}. Such two-component Bose-Hubbard models are also of interest in systems of ultracold atoms in optical traps.\cite{altman_phase_2003,han_quantum_2004} In that context, the two components correspond either to different species of atoms, or to two internal states of one atom species. We note that for cold atoms, the two-component Bose-Hubbard model has mainly been studied in the limit of negligible inter-component conversion, i.e.\ in our notation $g=0$. The relevance of atom-photon coupling in the JC lattice model serves as a motivation to explore the physics of the two-component Bose-Hubbard model in the regime of $g\not=0$ and large repulsion $U_\infty$ for one of the two components. We note that the hopping-dominated regime is easily obtained from \eqref{Uinf} by treating  inter-component conversion perturbatively, such that atoms and photons effectively decouple. This results in a superfluid of photons only weakly dressed by atomic excitations.

\subsection{Polariton mapping\label{sec:polaritonmodel}} 
The polariton mapping of the JC lattice model was previously discussed by Angelakis et al.\cite{angelakis_photon-blockade-induced_2007} In that work, the mapping was considered for the situation close to resonance, $\Delta\simeq0$. Here, we extend the mapping to the general case of nonzero detuning.

The idea of the polariton mapping is to diagonalize the local Jaynes-Cummings Hamiltonian $H^\text{JC}_j$ for each site $j$, and to rewrite the remaining photon hopping $H^\text{hop}$ in the form of polariton hopping between nearest neighbor sites.
Hence, the starting point for this mapping is the atomic limit, where the system's eigenstates are dressed-state polaritons of the form
\be
\ket{n,\pm} = a_{n\pm}\ket{n\downarrow} + b_{n\pm}\ket{(n-1)\uparrow}.
\ee
For $n\ge1$ the coefficients read
\begin{align}
a_{n\alpha}=\begin{cases} \sin \theta_n, & \alpha=+\\
                          \cos \theta_n, & \alpha=-
                          \end{cases},\quad
b_{n\alpha}=\begin{cases} \cos \theta_n, & \alpha=+\\
                          -\sin \theta_n, & \alpha=-
                          \end{cases},
\end{align}
while for the nondegenerate ground state $n=0$ we introduce the  conventions $\ket{0-}\equiv\ket{0\downarrow}$ and $\ket{0+}=0$, i.e.\ $a_{0-}=1$, $a_{0+}=b_{0\pm}=0$. (For the definition of the mixing angles $\theta_n$ see Appendix \ref{app:JC}.) With this, the linear relations between the basis states may be written compactly as 
\begin{align}
\left(
\begin{array}{l}
\ket{n+}\\
\ket{n-}
\end{array}
\right) &=
\left(
\begin{array}{ll}
	a_{n+} & b_{n+}\\
	a_{n-} & b_{n-}
\end{array}
\right)
\left(
\begin{array}{l}
	\ket{n\downarrow}\\
	\ket{(n-1)\uparrow}
\end{array}
\right),
\end{align}
with inverse
\begin{align}
\left(
\begin{array}{l}
\ket{n\downarrow}\\
\ket{(n-1)\uparrow}
\end{array}
\right) &=
\left(
\begin{array}{ll}
	a_{n+} & a_{n-}\\
	b_{n+} & b_{n-}
\end{array}
\right)
\left(
\begin{array}{l}
	\ket{n+}\\
	\ket{n-}
\end{array}
\right).
\end{align}
As required, the JC part of the Hamiltonian is now diagonal,
\be
\sum_j H^\text{JC}_j - \mu N = \sum_j \sum_{n=0}^\infty \sum_{\alpha=\pm}E^\mu_{n\alpha}P^\dag_{j,n\alpha}P_{j,n\alpha}.
\ee
Here, we have introduced the polariton operators in the same way as defined in Ref.\ \onlinecite{angelakis_photon-blockade-induced_2007}, namely
\be
P^\dag_{j,n\alpha}\equiv \ket{n\alpha}_j\bra{0-}_j
\ee
and their Hermitian conjugates. The operator $P^\dag_{j,n\alpha}$ maps the polariton ground state $\ket{0-}_j$ on site $j$ to another polariton state $\ket{n\alpha}_j$ on the same site.  Note that these operators do \emph{not} satisfy the canonical commutation relations of creation and annihilation operators. 

To rewrite the hopping contribution in the polariton basis, we observe that the effect of a photon annihilation operator can be expressed in the polariton basis via
\begin{align}
&a_j  \left(
\begin{array}{l}
	\ket{n+}_j\\
	\ket{n-}_j
\end{array}
\right)
=a_j\left(
\begin{array}{ll}
	a_{n+} & b_{n+}\\
	a_{n-} & b_{n-}
\end{array}
\right)
\left(
\begin{array}{l}
	\ket{ng}\\
	\ket{n-1,e}
\end{array}
\right)
\\
&=
\left(
\begin{array}{ll}
	t_{n,++} & t_{n,+-}\\
	t_{n,-+} & t_{n,--}
\end{array}
\right)
\left(
\begin{array}{l}
	\ket{n-1,+}\\
	\ket{n-1,-}
\end{array}
\right),
\end{align}
where the conversion amplitudes are  
\begin{align}
t_{n,\pm +} &= \sqrt{n}a_{n\pm} a_{n-1,+} +\sqrt{n-1} b_{n\pm} a_{n-1,-},\\
t_{n,\pm -} &= \sqrt{n}a_{n\pm} b_{n-1,+} +\sqrt{n-1} b_{n\pm} b_{n-1,-}.
\end{align}
As a result, the representation of the photon annihilation operator in the polariton basis reads
\begin{align}
a_j 
&= \sum_{n=1}^\infty \sum_{\alpha,\alpha'=\pm} t_{n,\alpha\alpha'} P_{j,(n-1)\alpha'}^\dag P_{j,n\alpha}
,
\end{align}
meaning that $a_j$ diminishes the polariton number $n$ on site $j$ by one, and may or may not alter the polariton type $\alpha=\pm$. 

In full, the JC lattice Hamiltonian can thus be rewritten as  
\begin{align}
H = &\sum_j \sum_{n=0}^\infty \sum_{\alpha=\pm} E^\mu_{n\alpha} P_{j,n\alpha}^\dag P_{j,n\alpha} \\\nonumber
&- \kappa \sum_j \sum_{j'\in\text{nn}(j)} \sum_{n,n'=1}^\infty\sum_{\alpha,\alpha',\beta,\beta'}t_{n,\alpha\alpha'}t_{n',\beta\beta'}\\\nonumber
&\qquad\qquad\times P^\dag_{j,n\alpha}P_{j,(n-1)\alpha'} P^\dag_{j',(n'-1)\beta'}P_{j',n'\beta},
\end{align}
where the polariton hopping indeed transfers polaritons between nearest-neighbor sites, and additionally may change the polariton type ($\pm$), of one or both sites involved.

It is instructive to compare this form of the JC lattice Hamiltonian to the Bose-Hubbard model. This comparison is facilitated by rewriting the Bose-Hubbard model $H=H_\text{BH}+H_\text{hop}$ in the corresponding form,
\begin{align}
H_\text{BH}&=\sum_j \varepsilon b_j^\dag b_j + \frac{1}{2}Un_j(n_j-1) = \sum_j \sum_{n=0}^\infty E_n P_{j,n}^\dag P_{j,n},\\
H_\text{hop}&= -\kappa \sum_{j}\sum_{j'\in\text{nn}(j)} b_j^\dag b_{j'} \\\nonumber
&= -\kappa \sum_{j}\sum_{j'\in\text{nn}(j)} \sum_{n,n'} \sqrt{n}\sqrt{n'} P_{j,n}^\dag P_{j,n-1} P_{j',n'-1}^\dag P_{j',n'},
\end{align}
with $E_n=n\varepsilon +U n(n-1)/2$. Inspection thus shows that the structures of the JC lattice Hamiltonian and the Bose-Hubbard Hamiltonian share several crucial properties. They both involve an onsite contribution and a hopping term which transfers excitations between nearest neighbor sites. The two system differ, however, in (i) the specific onsite eigenenergies, (ii) the scaling of the hopping matrix elements with $n$, and (iii) in the case of the JC lattice model, the existence of two excitation species and the possibility of interconversion.

\subsection{Mapping to the spin-1/2 quantum--XX model}
Angelakis et al.\ have pointed out that the interconversion between the two  polariton types may be neglected within RWA.\cite{angelakis_photon-blockade-induced_2007} Specifically,  their argument is based on the restriction to the subspace consisting of ground state and one-excitation manifold, and results in a mapping to the XX model.\footnote{Angelakis et al.\ choose to call the model given by Eq.\ \eqref{XXmodel} an XY model. Here we rather follow the nomenclature used, e.g., by Sachdev.\cite{sachdev_quantum_2000}} We review this statement, specify its range of validity and generalize it to include higher-excitation  manifolds.

Consider the regime close to a degeneracy point between two Mott-insulating lobes (polariton occupation numbers $n-1$ and $n$) in the atomic limit, i.e.\ $\kappa/g\ll1$. The effective model for low-energy states in this regime treats the two relevant product states $\ket{(n-1)-}^{\otimes_j}$ and $\ket{n-}^{\otimes_j}$. Given the conditions
\be
\kappa\ll\min\{|E^\mu_{(n+1)-}-E^\mu_{n-}|,|E^\mu_{(n-1)-}-E^\mu_{(n-2)-}|\}
\ee
and 
\be
\kappa\ll (E^\mu_{(n-1)+}-E^\mu_{(n-1)-}),
\ee
the hopping of photons will not lead to significant occupation of the symmetric Jaynes-Cummings states (higher in energy) or to occupation of adjacent Mott-insulating states. Thus, we may truncate the Hilbert space and drop inter-species conversion, which results in 
\begin{align}
H_{n}^\text{eff}=& \sum_j\sum_{m=-1,0} E^\mu_{(n+m)-} P_{j,(n+m)-}^\dag P_{j,(n+m)-} 
\\\nonumber
&- \kappa\,t_n^2 \sum_j\sum_{j'\in\text{nn}(j)} P^\dag_{j,n-}P_{j,(n-1)-} P^\dag_{j',(n-1)-}P_{j',n-}.
\end{align}
Note that the onsite energies $E^\mu_{(n-1)-}$, $E^\mu_{n-}$ and the  hopping amplitude $\kappa t_n\equiv \kappa t_{n,--}$ depend on the detuning $\Delta$ between resonator frequency and qubit frequency. Changing the detuning can thus be used to effectively alter the hopping strength, which provides a convenient experimental ``knob" to tune the system across the phase transition.\cite{angelakis_photon-blockade-induced_2007}

Within the truncated Hilbert space, the effective Hamiltonian $H^\text{eff}_n$ can be rewritten as a spin-lattice XX model. To see this, one identifies the polariton operators with spin-1/2 operators in the following way:
\be
\sigma_j^+ \equiv P_{j,n-}^\dag P_{j,(n-1)-} .
\ee
With this definition, $\sigma_j^+$ and its Hermitian conjugate $\sigma_j^-$ obey the standard commutation and anticommutation rules for the Pauli lowering and raising matrices. Specifically, we have
\begin{align}
[\sigma_j^+,\sigma_{j'}^-]&=\bigg[\ket{n-}_j\bra{(n-1)-}_j,\ket{(n-1)-}_{j'}\bra{n-}_{j'}\bigg]\nonumber\\
&=\delta_{jj'}\openone,\label{sigma1}
\end{align}
where we have used the completeness relation
\be \openone=\ket{n-}_j\bra{n-}_j+\ket{(n-1)-}_j\bra{(n-1)-}_j
\ee 
valid in the truncated Hilbert space of site $j$. Similarly, we confirm that
\be
\{ \sigma_j^+,\sigma_j^+\}=0, \quad \{ \sigma_j^+,\sigma_j^-\}=\openone.\label{sigma2}
\ee
Together with the usual definitions 
\be
\sigma_j^x=\sigma_j^++\sigma_j^-,\quad \sigma_j^y=-i\sigma_j^++i\sigma_j^-,\quad\sigma_j^z=2\sigma_j^+\sigma_j^--1,
\ee
Equations \eqref{sigma1} and \eqref{sigma2} are sufficient to reproduce the common algebra of Pauli spin matrices, i.e.\ for $a,b,c\in\{x,y,z\}$
\be
[\sigma^a_j,\sigma^b_j]=\delta_{jj'}2i\varepsilon_{abc}\sigma^c_j, \quad
\{\sigma^a_j,\sigma^b_j\}=2\delta_{ab}\openone.
\ee

 The resulting XX model describing the physics close to the degeneracy points of the JC lattice model is given by
\begin{align}
H^\text{XX}_n=&\frac{1}{2}\left(E^\mu_{n-}-E^\mu_{(n-1)-}\right)\sum_j\sigma^z_j \\\nonumber
&- \frac{1}{2}\kappa t_n^2 \sum_{\langle j,j' \rangle}\left(\sigma^x_j\sigma^x_{j'}+\sigma^y_j\sigma^y_{j'}\right)\label{XXmodel}.
\end{align}
This effective model underlines the similarity between the JC lattice model and the Bose-Hubbard model in the hardcore boson limit $U/\kappa\to\infty$, for which a similar XX model is obtained at the degeneracy points of adjacent Mott lobes.\cite{sachdev_quantum_2000}

\section{Effective field theory and existence of multicritical points\label{sec:fieldtheory}}
We now turn to the field-theoretic treatment of the JC lattice model.
Employing imaginary time coherent-state functional integration,  the partition function of the JC lattice system may be expressed as
\begin{align}
Z=\int \prod_j \mathcal{D}a_j^*(\tau)\mathcal{D}a_j(\tau)\mathcal{D}\vc{N}_j(\tau)\delta(\vc{N}_j^2-1)e^{-S[a_j^*,a_j,\vc{N}_j]},
\end{align}
where the action is given by
\begin{align}
S[a_j^*,a_j,\vc{N}_j]=&S_B  +\int_0^\beta d\tau\,\bigg\{\sum_j a_j^*\frac{\partial a_j}{\partial\tau}\\\nonumber
&+\sum_j H^\text{JC}_j(a_j^*,a_j,\vc{N}_j) +H^\text{hop}(a_j^*,a_j) \bigg\}.
\label{funcint}
\end{align}
Here, $H^\text{JC}$ and $H^\text{hop}$ are now the Hamilton functions obtained from the Hamiltonians, Eqs.\ \eqref{HJC2} and \eqref{Hhop}, by replacing all operators with their associated fields $a_j$, $a_j^*$,  and Bloch vectors $\vc{N}_j=(N_{j,x},N_{j,y},N_{j,z})$, i.e.\
\begin{equation}
a_j \to a_j(\tau), \, a_j^\dag \to a_j^*(\tau), \,\sigma_j^\alpha \to N_{j,\alpha},
\end{equation}
where $\alpha=x,y,z$. For convenience, we have absorbed the chemical potential term into the JC Hamiltonian, $H^\text{JC}_j\to H^\text{JC}_j-\mu n_j$. Further, the Berry phase contribution to the action from the spins is given by
\begin{equation}
S_B=\sum_j\int_0^\beta d\tau\, \langle \vc{N}_j(\tau) | \frac{d}{d\tau} | \vc{N}_j(\tau)\rangle .
\end{equation} 
(For a comprehensive exposition of coherent state functional integration of bosons and spins, see e.g.\ Refs.\ \onlinecite{sachdev_quantum_2000} and \onlinecite{negele_quantum_1998}.)

Analogous to the treatment of the Bose-Hubbard model,\cite{sachdev_quantum_2000,fisher_boson_1989} we next use a Hubbard-Stratonovich trans\-for\-ma\-tion to decouple the hopping term,
\begin{align}
&\exp\bigg[ \int_0^\beta d\tau\,\sum_{j,j'} a_j^*\kappa_{jj'}a_{j'}\bigg] \\\nonumber
&=\int \prod_j \mathcal{D}\psi_j^*(\tau)\psi_j(\tau)\exp\bigg[ -\int_0^\beta d\tau\, \sum_{j,j'}\psi_j^* \kappa_{jj'}^{-1} \psi_{j'}\bigg]\\\nonumber
&\qquad \times\exp\bigg[ \int_0^\beta d\tau\, \sum_j\left\{\psi_j^* a_j +\psi_j a_j^* \right\}\bigg].
\end{align}
Here, the nearest-neighbor hopping has been encoded in the matrix $\kappa_{jj'}$, yielding the photon hopping rate $\kappa$ if $j$ and $j'$ are nearest-neighbor sites, and zero otherwise. The introduction of the auxiliary fields $\psi_j$ thus results in an effective action that is completely local. In full, the partition function now reads
\begin{align}
Z=&\int\prod_j \mathcal{D}\psi_j^*(\tau)\mathcal{D}\psi_j(\tau)\mathcal{D}a_j^*(\tau)\mathcal{D}a_j(\tau)\mathcal{D}\vc{N}_j(\tau)\delta(\vc{N}_j^2-1)\nonumber\\
&\times\exp\bigg(- S'[\psi_j^*,\psi_j,a_j^*,a_j,\vc{N}_j]\bigg)\label{partitionf},
\end{align}
with an action $S'$ that includes the auxiliary fields $\psi_j^*$, $\psi_j$:
\begin{align}\label{sprime}
&S'=S_B +\int_0^\beta d\tau\,\bigg\{\sum_j a_j^*\frac{\partial a_j}{\partial\tau} +\sum_{j,j'}\psi_j^*\kappa_{jj'}^{-1}\psi_{j'} \\\nonumber
&\qquad+\sum_j H^\text{JC}_j(a_j^*,a_j,\vc{N}_j) -\sum_j(\psi_j^*a_j + \psi_j a_j^*)\bigg\}.
\end{align}
In the next step, we integrate out the fields $a_j^*$, $a_j$, and $\vc{N}_j$. For the discussion of the quantum phase transition,  we are primarily interested in an effective field theory for the auxiliary fields $\psi_j^*$ and $\psi_j$, which should hold for ``small fields" $\psi$. (The fields $\psi_j$ are proportional to the order parameter $\langle a_j \rangle$, and hence are small in the critical region.) Thus, we may expand in the fields $\psi_j$ and their temporal and spatial gradients to obtain the general effective action
\begin{equation}\label{seff}
S_\text{eff}[\psi^*,\psi]=\int_0^\beta d\tau\,\int d^dx\,\mathcal{L}_\text{eff}(\psi^*(x,\tau),\psi(x,\tau)),
\end{equation}
with Lagrangian density 
\begin{align}\label{Leff}
\mathcal{L}_\text{eff}= &K_0 + K_1\psi^*\frac{\partial\psi}{\partial\tau} + K_2\abs{\frac{\partial\psi}{\partial\tau}}^2
+ K_3 \abs{\nabla \psi}^2 \\\nonumber
&+ \tilde{r}\abs{\psi}^2 + \frac{\tilde{u}}{2}\abs{\psi}^4+\cdots
\end{align}
Next, we wish to obtain explicit expressions for the coefficients in this expansion. As in the Bose-Hubbard model,\cite{sachdev_quantum_2000,herbut_modern_2007} the coefficients $\tilde{r}$ and $\tilde{u}$ can be related to the mean-field coefficients $r$ and $u$, Eq.\ \eqref{mfexp}. To see this, consider the situation of a spatially and temporally constant field $\psi_j(\tau)=\psi$. In this case, we identify 
\begin{align}\nonumber
&\int\prod_j \mathcal{D}a_j^*(\tau)\mathcal{D}a_j(\tau)\mathcal{D}\vc{N}_j(\tau)\delta(\vc{N}_j^2-1)e^{-S'[a_j^*,a_j,\vc{N}_j;\psi]}\\
&\simeq e^{-S_\text{eff}(\psi)}\equiv Z(\psi),
\end{align}
where the effective action \eqref{seff} reduces to
\begin{equation}
S_\text{eff}(\psi)=\beta V \left( K_0 + \tilde{r}\abs{\psi}^2 + \frac{u}{2}\abs{\psi}^4+\cdots \right).
\end{equation}
Here, $V$ denotes the volume of the system. The substitution of the global field $\psi$ into the action $S'$, Eq.\ \eqref{sprime}, is facilitated by the relation $\sum_{ij}\kappa_{ij}^{-1}=\mathcal{N}/(\kappa z_c)$ where $\mathcal{N}$ is the total number of sites. By inspection, one can verify that $Z(\psi)$ is the partition function corresponding to the mean-field Hamiltonian
\begin{align}
&H_\text{mf} = \sum_j\bigg[ \frac{1}{2}(\varepsilon-\mu)\sigma_j^z + (\omega-\mu) a_j^\dag a_j \nonumber\\
&\quad +  g(a_j^\dag \sigma_j^- + \sigma_j^+ a_j)  - (a_j \psi^* + a_j^\dag \psi) +\frac{1}{z_c\kappa}\abs{\psi}^2\bigg].
\end{align}
In the limit of low temperature, $\beta\to\infty$, and close to the phase boundary (i.e.\ small $\psi$), we thus identify
\begin{equation}
S_\text{eff}(\psi)=\beta E_0(\psi),
\end{equation}
where $E_0(\psi)$ is the ground-state energy of the mean-field Hamiltonian $H_\text{mf}$. As a result, we conclude 
\begin{equation}
\tilde{r}=v^{-1}r,\qquad \tilde{u}=v^{-1}u,
\end{equation}
where $v=V/\mathcal{N}$ is the volume per lattice site.

To obtain the coefficients $K_1$ and $K_2$, we generalize the time-dependent U(1) symmetry encountered for the Bose-Hubbard model\cite{sachdev_quantum_2000} to the JC lattice model. In the case of the latter model, the symmetry involves both the photon fields as well as the spins:
\begin{align}\nonumber
a_j&\to a_j e^{i\varphi(\tau)},\\\label{gauget}
\vc{N}_j&\to\left(
\begin{array}{rrr}
\cos\varphi & \sin\varphi & 0\\
-\sin\varphi & \cos\varphi & 0\\
0 & 0 & 1	
\end{array}
\right)\vc{N}_j,\\\nonumber
\psi_j &\to \psi_j e^{i\varphi(\tau)},\\\nonumber 
(\omega-\mu) &\to (\omega -\mu) -i \frac{d\varphi}{d \tau}.
\end{align}
This symmetry transformation leaves the action $S'$, Eq.\ \eqref{sprime} invariant.\footnote{We note that, as in the Bose-Hubbard case, the gauge transformation Eq.\ \eqref{gauget} actually leads out of the physical space, since $\omega-\mu$ acquires an imaginary part.} The only two terms for which invariance may not be immediately obvious are the atom-photon coupling term and the spin Berry phase term $S_B$. The verification of invariance for these two terms is given in Appendix \ref{app:invariance}.

With the invariance intact, we now explore its consequences for the expansion coefficients in Eq.\ \eqref{Leff}. Specifically, we consider the transformation in the limit of a small phase $\varphi(\tau)\ll 1$, and plug the transformed fields $\psi^*$, $\psi$, and the transformed $\omega-\mu$ into the effective action $S_\text{eff}[\psi^*,\psi]$. Taking into account that the coefficients $\tilde{r}$, $\tilde{u}$, and $K_\ell$  may depend on $\omega-\mu$, we Taylor-expand in powers of the phase $\varphi$ and its time derivatives.  Requiring that $S_\text{eff}$ be invariant, the leading order in this expansion yields the important relations
\begin{equation}
K_1=\frac{\partial \tilde{r}}{\partial (\omega-\mu)},\qquad K_2=-\frac{1}{2}\frac{\partial^2 \tilde{r}}{\partial (\omega-\mu)^2},\label{Keq}
\end{equation}
see Appendix \ref{app} for a more detailed derivation. 

Whenever the coefficient $K_1$ vanishes, the phase transition of the JC lattice system changes its universality class. While the case of $K_1\not=0$ is associated with the generic superfluid-Mott insulator transition with dynamical critical exponent $z=2$, points where $K_1=0$ lead to a phase transition within the universality class of the soft-spin $(d+1)$--dimensional XY model. The physics of these multicritical points is in complete analogy to the corresponding physics of the Bose-Hubbard model.\cite{fisher_boson_1989,sachdev_quantum_2000} 

\begin{figure}
\centering
		\includegraphics[width=0.7\columnwidth]{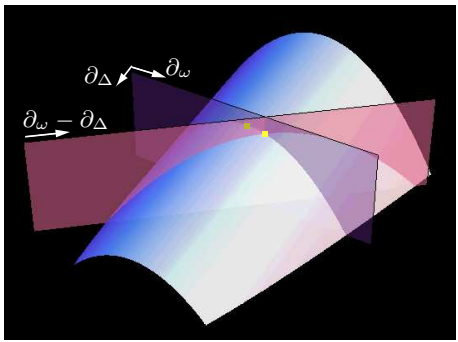}
	\caption{(Color online) Change of variables shifts multicritical points away from lobe tips. While multicritical points appear at lobe tips when plotting the phase boundary as a function of $(\omega-\mu)$ and $\kappa$ for constant atomic energy $\varepsilon$, their position is shifted when plotting the phase boundary for constant detuning $\Delta$, see Eq.\ \eqref{derivs}.
	\label{fig:deriv}}
\end{figure}

One difference worth noting is the presence of an additional parameter in the case of the JC lattice system. This parameter, the energy scale for the atoms $\varepsilon$, is to be held constant when calculating the partial derivative in Eq.\ \eqref{Keq}. Due to the presence of this additional parameter, the phase boundary is a two-dimensional surface in the space spanned by the parameters $\omega$, $\varepsilon$, and $\kappa$, and the condition $K_1=0$ defines curves on the phase boundary. Consequently, the field theory predicts a multicritical \emph{curve}, as opposed to the isolated multicritical points of the Bose-Hubbard model.  

These curves are shown in Fig.\ \ref{fig:3dphase}, where the axes have been chosen as resonator frequency $\omega$ and detuning $\Delta$. As mentioned by Zhao et al.,\cite{zhao_insulator_2008} the transformation of variables $\{\omega,\varepsilon\}\to\{\omega,\Delta\}$ shifts the position of multicritical points away from the lobe tips, when considering cuts for constant detuning $\Delta$, compare Fig.\ \ref{fig:deriv}. Specifically, the change of variables implies the transformation of the partial $\omega$ derivative according to 
\be\label{derivs}
\frac{\partial}{\partial\omega}\bigg|_{\varepsilon}=\frac{\partial}{\partial\omega}\bigg|_{\Delta}-\frac{\partial}{\partial\Delta}\bigg|_{\omega},
\ee
where subscripts to the partial derivatives denote which variable to hold constant.

We thus predict \emph{multicritical curves}, whose position is completely specified through the conditions given by Eq.\ \eqref{pboundary}\footnote{While the phase boundary, Eq.\ \eqref{pboundary}, is obtained within mean-field approximation and may acquire corrections when including the effects of fluctuations, see e.g. Ref.\ \onlinecite{schmidt_strong_2009}, we emphasize that the Equations \eqref{Keq} and \eqref{derivs} are exact.}
\be
\frac{\partial}{\partial\omega}\bigg|_{\varepsilon}R_n = 0.
\ee
For each Mott lobe with $n=1,2,\ldots$, there is one such multicritical curve, along which the universality class changes in a way similar to the Bose-Hubbard model. It is interesting to note that the numerical evidence for these multicritical curves is currently discussed controversially. Specifically, Zhao et al.\ argue for the complete absence of multicritical points based on their quantum Monte Carlo simulations.\cite{zhao_insulator_2008} By contrast, Schmidt and Blatter have recently presented new evidence for the presence of such multicritical points,\cite{schmidt_strong_2009} which is consistent with our findings from field theory. While the exact position and shape of the phase boundary may acquire corrections beyond the mean-field results presented in Eqs.\ \eqref{Req}--\eqref{pboundary}, we emphasize that the field-theoretical analysis is general and applies to arbitrary spatial dimensions of the system. Specifically, as long as the phase boundary remains differentiable (i.e.\ no kinks) it predicts the existence of multicritical lines independent of the spatial dimension of the lattice.

\section{Conclusions and Outlook\label{sec:conclusions}}
In conclusion, we have presented a thorough analysis of the quantum phase transition predicted for polaritons in the Jaynes-Cummings lattice. We have revisited the mean-field phase diagram, clarified the existence of an unstable region, and derived analytical expressions for the full two-dimensional surface that constitutes the phase boundary. The qualitative similarities between the Jaynes-Cummings lattice and the Bose-Hubbard model have been elaborated and discussed in the context of exact mappings to a two-component Bose-Hubbard model and a generalized polariton model. Finally, we have presented a field theoretic approach for the Jaynes-Cummings lattice, whose gradient expansion in the critical region underlines the analogy with the Bose-Hubbard model. We find that the Jaynes-Cummings lattice falls in the same universality class, and have proven the existence of multicritical curves which parallel the presence of multicritical points in the Bose-Hubbard model.

\begin{figure}
\centering
		\includegraphics[width=1.0\columnwidth]{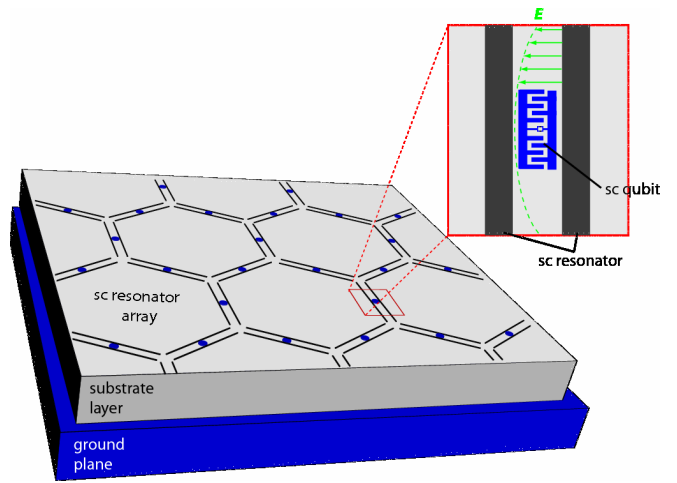}
	\caption{(Color online) Possible realization of the JC lattice model, building upon circuit quantum electrodynamics: superconducting microwave resonators can be patterned onto a chip and coupled to superconducting qubits (shown in the zoom-in). \label{fig:exp}}
\end{figure}
A number of potential candidates have been suggested for an actual realization of the Jaynes-Cummings lattice physics, ranging from arrays of photonic band-gap cavities to realizations in circuit QED; see the review by Hartmann et al.\ and references therein.\cite{hartmann_quantum_2008-1} The realization in circuit QED, see Fig. \ref{fig:exp}, is particularly interesting. In this case, the fabrication of the basic building block is well established and shows the desired Jaynes-Cummings physics.\cite{wallra_strong_2004,schoelkopf_wiring_2008} The step to fabricating medium-size arrays ($\mathcal{N}\sim10^1$--$10^2$) should not pose fundamental difficulties, and first ideas how to initialize such a system have been suggested.\cite{angelakis_photon-blockade-induced_2007} Namely, the system could be prepared within a  Mott-insulating phase by using a global external microwave signal. An off-resonant ac Stark tone could then be used to tune the system across the phase boundary by modifying the detuning $\Delta$, and would also allow for the study of the system at multicritical points. By contrast, both $\kappa$ as well as $\omega$ would typically be fixed by the fabrication parameters. The order parameter $\psi=\langle a\rangle$  could be accessed by a homodyne measurement of the voltage field in the resonators. 

In the future, these ideas will have to be extended to specify actual experimental protocols. As pointed out previously,\cite{rossini_mott-insulating_2007} an analysis of the Jaynes-Cummings lattice when subject to unavoidable decoherence mechanisms will be crucial to aid the experimental attempt. An additional layer of interesting physics will arise from disorder. Specifically, in the circuit QED architecture such disorder is likely to occur for the atoms' frequencies due to the limited precision in Josephson junction fabrication.    We are confident that a concerted effort of theory and experiment can lead to the realization and further study of Jaynes-Cummings lattice physics in the near future, and note that such a realization would share many of the interesting aspects and benefits of the thriving field of ultracold atom physics.

\begin{acknowledgments}
This work was supported by the NSF through the Yale Center for Quantum Information Physics (DMR-0653377).
\end{acknowledgments}

\appendix
\section{Solution of the JC model\label{app:JC}}
For completeness, we compile the most important results for the Jaynes-Cummings model.
The JC Hamiltonian,\cite{jaynes_comparison_1963}
\begin{equation}
H^\text{JC} =  \omega a^\dag a + \varepsilon\sigma^+\sigma^- +g(a^\dag\sigma^- + \sigma^+a),
\end{equation}
describes the situation of a two-level system (transition energy $\varepsilon$, Pauli raising and lowering operators $\sigma^{\pm}$) coupled to a harmonic oscillator (frequency $\omega$, creation and annihilation operators $a^\dag$, $a$) with coupling strength $g$. Prominent realizations of Jaynes-Cummings physics include cavity and circuit QED,\cite{raimond_colloquium:_2001,schoelkopf_wiring_2008} where the JC Hamiltonian is usually obtained within the rotating wave approximation (RWA) valid for $g\ll\omega,\varepsilon$.

The RWA is crucial: within this approximation, the total number of excitations $N=a^\dag a+\sigma^+\sigma^-$ is conserved and the Hamiltonian separates into  $2\times2$--blocks for each fixed-excitation subspace, i.e.\ $H^\text{JC}=\bigotimes_{n=0}^\infty h_n$, where $h_0=0$ and
\begin{equation}
h_n=\left(
\begin{array}{ll}
	 n\omega & \sqrt{n}g\\
	 \sqrt{n}g & (n-1)\omega +\varepsilon
\end{array} \right).
\end{equation}
 Precisely because of this block-diagonal form, the JC model is exactly solvable, requiring merely the diagonalization of a $2\times2$--matrix. The resulting energy spectrum is given by the ground state energy $E_0=0$ and the excited state energies ($n\ge1$)
 \begin{equation}
 E_{n\pm}=n\omega+\frac{\Delta}{2}\pm\bigg[\left(\frac{\Delta}{2}\right)^2+n g^2\bigg]^{1/2}.
 \end{equation}
 Here, we have used $\Delta=\varepsilon-\omega$ as an abbreviation for the detuning, and the polariton number is counted by $n\in\NN$. The eigenstates corresponding to these eigenenergies are $\ket{0\downarrow}$ for the ground state and
 \begin{align}
 \ket{n-}&=\sin\theta_n \ket{n\downarrow} + \cos\theta_n \ket{(n-1)\uparrow},\\
 \ket{n+}&=\cos\theta_n \ket{n\downarrow} - \sin\theta_n \ket{(n-1)\uparrow}. 
 \end{align}
 The mixing angle $\theta_n$ depends on detuning, coupling strength, and polariton number:
 \begin{equation}
 \theta_n = \frac{1}{2}\arctan\left( \frac{2g\sqrt{n}}{\Delta} \right).
 \end{equation}

\section{Perturbative shifts for $\Delta\not=0$\label{app:pert}}
In the critical region and within mean-field theory, the energy shifts $R_n\abs{\psi}^2$ due to photon hopping can be calculated via perturbation theory. The resulting expressions for vanishing detuning were given in Section \ref{sec:meanfield}. Here, we provide the slightly lengthier but general expressions:
\begin{align}
&R_0 =-\bigg[\frac{\cos^2 \theta_1}{E^\mu_{1-}}+\frac{\sin^2 \theta_1}{E^\mu_{1+}}\bigg]\\
&R_n = \\\nonumber
&=\bigg[\frac{\abs{\sqrt{n+1}\cos\theta_n\cos\theta_{n+1}+\sqrt{n}\sin\theta_n\sin\theta_{n+1}}^2}{E^\mu_{n-}-E^\mu_{(n+1)-}}\\\nonumber
&\quad+ 
\frac{\abs{\sqrt{n+1}\cos\theta_n\sin\theta_{n+1}-\sqrt{n}\sin\theta_n\cos\theta_{n+1}}^2}{E^\mu_{n-}-E^\mu_{(n+1)+}}\\\nonumber
&\quad+
\frac{\abs{\sqrt{n}\cos\theta_n\cos\theta_{n-1}+\sqrt{n-1}\sin\theta_n\sin\theta_{n-1}}^2}{E^\mu_{n-}-E^\mu_{(n-1)-}}\\\nonumber
&\quad+
\frac{\abs{\sqrt{n}\cos\theta_n\sin\theta_{n-1}-\sqrt{n-1}\sin\theta_n\cos\theta_{n-1}}^2}{E^\mu_{n-}-E^\mu_{(n-1)+}}\bigg].
\end{align}
(See the previous Appendix \ref{app:JC} for the definition of the mixing angle $\theta_n$.)

\section{Invariance of atom-photon coupling and spin Berry phase under local U(1) gauge transformation\label{app:invariance}}
In Section \ref{sec:fieldtheory}, we have claimed that the action $S'$ given in Eq.\ \eqref{sprime} is invariant with respect to the local U(1) gauge transformation specified in Eq.\ \eqref{gauget}. The invariance is verified by simple inspection. For completeness, we briefly discuss the proof for the atom-photon coupling and the spin Berry phase term.

First, consider the coupling term
\begin{equation}
g(a_j^*[N_{j,x}-iN_{j,y}] +\text{c.c.})
\end{equation}
This is transformed into
\begin{align}\nonumber
&g\bigg(a_j^*e^{-i\varphi}[N'_{j,x}-iN'_{j,y}] +\text{c.c.}\bigg)=
g\bigg(a_j^*e^{-i\varphi}[\cos\varphi N_{j,x} \\\nonumber
&\qquad\qquad + \sin\varphi N_{j,y} +i \sin\varphi N_{j,x} -i\cos\varphi N_{j,y}] +\text{c.c.}\bigg)\\\nonumber
&=g\bigg(a_j^*e^{-i\varphi}[e^{i\varphi}N_{j,x}-i e^{i\varphi} N_{j,y}]+\text{c.c}\bigg)\\
&=g\bigg(a_j^*[N_{j,x}-iN_{j,y}] +\text{c.c.}\bigg),
\end{align}
which reveals the invariance.
Second, consider the transformation of $S_B$. Writing the rotation matrix affecting $\vc{N}_j$ as $\mathsf{R}(\tau)$, we have
\begin{equation}
S_B\to S_B + \sum_j\int_0^\beta d\tau\,\langle \vc{N}_j(\tau) | \mathsf{R}^\dag(\tau)\dot{\mathsf{R}}(\tau)|\vc{N}_j(\tau)\rangle.
\end{equation}
Evaluating the matrix derivative and multiplying with the transpose of $\mathsf{R}$, we find
\begin{equation}
\mathsf{R}^\dag(\tau)\dot{\mathsf{R}}(\tau)=\frac{d\varphi}{d\tau}\left( 
\begin{array}{crr}
0 & 1 & 0\\
-1 & 0 & 0\\
0 & 0& 0	
\end{array}
\right),
\end{equation}
which gives $\langle \vc{N}_j(\tau) | \mathsf{R}^\dag(\tau)\dot{\mathsf{R}}(\tau)|\vc{N}_j(\tau)\rangle=0$, and thus proves the invariance of $S_B$.

\section{Derivation of relations between $K_1$, $K_2$, and $\tilde{r}$\label{app}}
Our derivation of the interrelations between the coefficients $K_1$, $K_2$, and $\tilde{r}$ which enter the effective action for the auxiliary field $\psi$, Eqs.\ \eqref{seff} and \eqref{Leff}, is analogous to the reasoning applied in the case of the Bose-Hubbard model; see, e.g., Refs.\ \onlinecite{sachdev_quantum_2000} and \onlinecite{herbut_modern_2007}. Starting point of this consideration is the invariance of the action $S'$, Eq.\ \eqref{sprime},
under the generalized gauge transformation specified in Eq.\ \eqref{gauget}. Once the Bose fields $a_j$, $a_j^*$, and the spins $\vc{N}_j$ have been integrated out, this invariance remains intact in the following sense: when transforming the auxiliary fields and the cavity frequency according to
\begin{equation}
\psi_j\to\psi_j e^{i\varphi(\tau)},\qquad (\omega-\mu)\to(\omega-\mu)-i\frac{d\varphi}{d\tau},
\end{equation}
the effective action $S'$
with Lagrangian density 
\begin{align}
\mathcal{L}_\text{eff}=& K_0 + \frac{1}{2}K_1\left( \psi^*\frac{\partial\psi}{\partial\tau}+\frac{\partial\psi^*}{\partial\tau}\psi\right) + K_2\abs{\frac{\partial\psi}{\partial\tau}}^2
\nonumber\\
& + K_3 \abs{\nabla \psi}^2 + \tilde{r}\abs{\psi}^2 + \frac{1}{2}\tilde{u}\abs{\psi}^4+\cdots
\end{align}
remains unchanged. This statement must hold for arbitrary configurations of the fields $\psi_j^*$ and $\psi_j$, as well as arbitrary choices of the phase function $\varphi(\tau)$. As a consequence, the invariance must not only hold for the effective action, but also for the Lagrangian density itself. In particular, the coefficients $K_1$, $K_2$ etc.\ themselves must be invariant, and they must be so to each order in $\varphi$, $\dot\varphi$, and its higher time derivatives, when considering a small global phase change $\varphi(\tau)\ll1$. 

Carrying out this expansion for the terms included in $\mathcal{L}_\text{eff}$ above, one finds
\begin{align}\label{L1}
\mathcal{L}_\text{eff}\simeq &K_0 -i\dot\varphi K_0'-\frac{1}{2}\dot\varphi^2 K_0''\\\nonumber
&+\left[\left(\frac{K_1}{2} -i\dot\varphi \frac{K_1'}{2} -i\dot\varphi K_2\right)\psi^*\frac{\partial\psi}{\partial\tau}+\text{c.c.}\right] \\\nonumber
&+(K_2 -i\dot\varphi K_2')\abs{\frac{\partial\psi}{\partial\tau}}^2
+(K_3 -i\dot\varphi K_3')\abs{\nabla\psi}^2\\\nonumber
&+(\tilde{r}-i\dot\varphi\tilde{r}'+i K_1\dot\varphi)\abs{\psi}^2+
\frac{1}{2}(\tilde{u}-i\dot\varphi \tilde{u}')\abs{\psi}^4 +\cdots,
\end{align}
where $\dot\varphi=d\varphi/d\tau$, and primes denote derivatives with respect to $(\omega-\mu)$. It is important to note that the terms collected in Eq.\ \eqref{L1} do \emph{not} yet correspond to a consistent expansion. Specifically, higher-order terms dropped in $\mathcal{L}_\text{eff}$ above give relevant contributions even to the order recorded in Eq.\ \eqref{L1}. As one example, consider the higher-order term
\begin{equation}
\alpha_1 \frac{\partial\psi^*}{\partial\tau}\frac{\partial^2\psi}{\partial\tau^2},
\end{equation}
whose expansion contains (among others!) the contribution
\begin{equation}
i \alpha_1 \dot\varphi \abs{\frac{\partial\psi}{\partial\tau}}^2.
\end{equation}
One can verify, however, that the coefficients of the $\abs{\psi}^2$ and the $\psi^*\partial\psi/\partial\tau$ terms \emph{are} in fact consistent to the order $\mathcal{O}(\dot\varphi)$ as written above, and we can read off the important relations
\begin{equation}
K_1=\tilde{r}',\qquad K_2=-\frac{1}{2}K_1'.
\end{equation}


\end{document}